\documentclass[oneside,12pt]{article}

\usepackage{amsmath,amsgen,amstext,amsbsy,amsopn,amsthm, amssymb}

\newtheorem{thm}{Theorem}
\numberwithin{thm}{section}

\numberwithin{cor}{section}
\newtheorem{lem}{Lemma}
\numberwithin{lem}{section}
\newtheorem{prop}{Proposition}
\numberwithin{prop}{section}
\theoremstyle{definition}

\numberwithin{defn}{section}
\newtheorem{rem}{Remark}
\numberwithin{rem}{section}

\newcommand{\D}{\mathcal{D}}
\newcommand{\pd}{\partial}
\newcommand{\eps}{\epsilon}
\newcommand{\suli}{\sum\limits}
\newcommand{\inli}{\int\limits}

\newcommand{\aob}{\left(\frac{a}{b}\right)}
\newcommand{\V}{V}
\newcommand{\al}{\alpha}
\newcommand{\xij}{|x_i-x_j|}
\newcommand{\half}{\mbox{$\frac{1}{2}$}}
\newcommand{\E}{\mathcal{E}}
\newcommand{\FF}{\mathcal{F}}
\newcommand{\F}{F}
\newcommand{\G}{G}
\newcommand{\rmax}{\rho_{\al,\text{max}}}
\newcommand{\rmin}{\rho_{\al,\text{min}}}
\newcommand{\R}{\mathbb{R}}
\newcommand{\real}{\text{Re}}
\newcommand{\imag}{\text{Im}}
\newcommand{\e}{\widetilde{e}}
\newcommand{\vv}{\widetilde{v}}
\newcommand{\RR}{\widetilde{R}}
\newcommand{\as}{\widetilde{a}}
\numberwithin{equation}{section}  

\begin{document}
\markboth{\scriptsize{LSY 05/08/99}}{\scriptsize{LSY 05/08/99}}
\title{\bf{Bosons in a Trap:\\A Rigorous Derivation of the\\
Gross-Pitaevskii Energy Functional}}
\author{\vspace{5pt} Elliott H.~Lieb$^1$, Robert Seiringer$^{2}$, and Jakob 
Yngvason$^{2}$\\ 
\vspace{-4pt}\small{$1.$ Departments of Physics and Mathematics, Jadwin Hall,} \\
\small{Princeton University, P.~O.~Box 708, Princeton, New Jersey
  08544}\\
\vspace{-4pt}\small{$2.$ Institut f\"ur Theoretische Physik, Universit\"at 
Wien}\\
\small{Boltzmanngasse 5, A 1090 Vienna, Austria}}
\date{\small 5 August, 1999}
\maketitle

\footnotetext[1]{Work partially
supported by U.S. National Science Foundation
grant PHY 98-20650.\\
\copyright\,1999 by the authors. This paper may be reproduced, in its
entirety, for non-commercial purposes.}

\begin{abstract}
The ground state properties of interacting Bose gases in external
potentials, as considered in recent experiments, are usually described
by means of the Gross-Pitaevskii energy functional. We present here the
first proof of the asymptotic exactness of this approximation for the
ground state energy and particle density of a dilute Bose gas with a
positive interaction.  \end{abstract}

\section{Introduction}

Recent experimental breakthroughs in the treatment of dilute Bose gases
have renewed interest in formulas for the ground state  and its energy
derived many decades ago. One of these is the Gross-Pitaevskii
(GP) formula for the energy in a trap \cite{G1961}--\cite{G1963}, such as is used in the actual experiments. We refer to \cite{DGPS} for an up to date review of this 
approximation and its applications.
One of the inputs needed for its justification is the
ground state energy per unit volume  of a dilute, thermodynamically
infinite, homogeneous gas. This latter quantity has been known for many
years, but it was only very recently that it was derived rigorously 
\cite{LY1998} for suitable interparticle potentials. Consequently, it
is appropriate now to use this new result to go one step further and
derive the GP formula rigorously.

The starting point for our investigation is the Hamiltonian for $N$ identical 
bosons
\begin{equation}\label{ham}
H^{(N)}=\suli_{i=1}^{N}\left(-\nabla_i^2+\V(x_i)\right) +\suli_{i<j}v(\xij)
\end{equation} 
acting on totally {\it symmetric}, square integrable wave
functions of $(x_1,\dots,x_N)$ with
$x_i\in\R^{3}$.  Units have here been chosen so that $\hbar=2m=1$, where 
$m$ is the mass. We consider external potentials $\V$ that are measurable and 
locally bounded and tend to infinity for $|x|\to\infty$ in 
the
sense that 
$\inf_{|x|\geq R}\V(x)\to\infty$ for
$R\to\infty$.  
The potential is then bounded below and for convenience we assume that its 
minimum value is zero. The ground state of
$-\nabla^2+\V(x)$ 
provides
a natural energy unit, $\hbar\omega$, and the 
corresponding length unit, $\sqrt{\hbar/m\omega}$, describes the extension of 
the potential. We shall measure all energies and lengths in these units.
In the available experiments $V$ is typically $\sim|x|^2$ and 
$\sqrt{\hbar/m\omega}$ 
of the order $10^{-6}$ m.

The
particle interaction $v$ is assumed to be positive, spherically symmetric
and 
decay faster than $|x|^{-3}$ at infinity.  In
particular, the scattering length, denoted by $a$, should be finite. We 
recall that the (two-body) scattering length is defined by means of 
the solution $u(r)$ of the zero energy scattering equation
\begin{equation}\label{scatteq}
-u''(r)+\mbox{${\frac 12}$} v(r)u(r)=0
\end{equation}
with $u(0)=0$; by definition, $a=\lim_{r\to\infty}(r-u(r)/u'(r))$. Let 
$v_1(r)$ be a fixed potential with scattering length $a_1$. Then 
$v(r)=(a_1/a)^2v_1(a_1r/a)$ has scattering length $a$. We regard in the 
following $v_1$ as fixed, but vary $a$ (in fact, we shall take $a=a_1/N$). The 
ground 
state energy $E^{\rm QM}$ of \eqref{ham} depends on the potentials $\V$ and 
$v$, 
besides $N$, but with $\V$ fixed and $v(r)=(a_1/a)^2v_1(a_1r/a)$, 
the notation $E^{\rm QM}(N,a)$ is justified. The corresponding
eigenfunction will be denoted $\Psi^{(N)}_0$.  It is
unique up to a phase that can be chosen such that the wave function is
strictly positive where the interaction is finite \cite{reed}.  
The
particle density is defined by 
\begin{equation}\label{qmdensity} 
\rho^{\rm QM}_{N,a}(x)=N\int_{\R^{3(N-1)}}
|\Psi^{(N)}_0(x,x_2,\dots,x_N)|^2d^3x_2\dots d^3x_N.  
\end{equation} 
\medskip

The {\bf Gross-Pitaevskii (GP) energy functional} is defined as
\begin{equation}\label{gpfunct}
\E^{\rm GP}[\Phi]=\int_{\R^3}\left(|\nabla\Phi(x)|^2+\V(x)|\Phi(x)|^2+4\pi 
a|\Phi(x)|^4\right)d^3x
\end{equation}
where $\Phi$ is a function on $\R^3$. For a given $N$ the corresponding GP 
energy, denoted  $E^{\rm GP}(N,a)$, is defined as the infimum of $\E[\Phi]$ 
under the normalization condition 
\begin{equation}\label{norm}\int_{\R^3}|\Phi(x)|^2d^3x=N.
\end{equation}  
It has the simple 
scaling property
\begin{equation}\label{scaling}
E^{\rm GP}(N,a)=NE^{\rm GP}(1,Na).
\end{equation}
What  \eqref{scaling} shows is that the GP functional \eqref{gpfunct} 
together with the normalization condition \eqref{norm} has one 
characteristic parameter, namely  $Na$. (Recall that lengths are 
measured in the unit $\sqrt{\hbar/m\omega}$ associated with $V$ so $a$ is 
dimensionless.) Thus, if we want to investigate the non-trivial 
aspects of GP theory we have to consider a limit in which 
$N\to\infty$ with $Na$ fixed.  This 
explains the seemingly peculiar limit in Theorems 1.1 and 1.2. As $Na\to\infty$
the GP energy functional simplifies, since the gradient term becomes
small compared to the other terms,
and the so called \lq\lq Thomas-Fermi 
limit\rq\rq\ described in Theorem \ref{TFlimit} results.
In some typical experiments $a$ is about $10^{-3}$, while $N$ varies 
from $10^3$ to $10^7$. Thus $a_{1}$ in Theorems 1.1 and 1.2 varies 
from 1 to about $10^4$.

In the next section it will be
shown that the infimum of the energy functional (\ref{gpfunct}), under the
subsidiary condition (\ref{norm}),  is obtained for a unique, 
strictly
positive function, denoted $\Phi^{\rm GP}$.  The GP density is given by
\begin{equation}\label{gpdensity} 
\rho_{N,a}^{\rm GP}(x)=\Phi^{\rm GP}(x)^2. \end{equation}
It satisfies
\begin{equation} \rho_{N,a}^{\rm GP}(x)= N\rho_{1,Na}^{\rm 
GP}(x).\label{rhoscaling}\end{equation}

The main result of this paper concerns the behavior of the quantum mechanical 
ground state energy $E^{\rm QM}(N,a)$ when $N$ is large, but $a$ is small, so 
that $Na$ is $O(1)$. 
It is important to note that although the density tends to infinity 
for $ N\to \infty$ (by Eq.\ (\ref{rhoscaling})) we are still concerned
with 
{\it dilute} systems in 
the sense that $a^3\bar \rho\ll 1$, where
\begin{equation}\label{rhobar}
\bar\rho=\frac 1N\int\rho^{\rm GP}_{N,a}(x)^2 d^3x
\end{equation}
is the {\bf mean GP density}. (Note the exponent 2 in
(\ref{rhobar}).) In fact, since 
$a\sim N^{-1}$, 
$a^3\bar\rho\sim N^{-2}$.

The precise statement of the limit theorem for the energy is as follows.

\begin{thm}[The GP energy is the dilute limit of the QM energy]\label{thm1}
For every fixed $a_1$
\begin{equation}\lim_{N\to\infty}\frac{E^{\rm QM}(N,a_1/N)}{N}=E^{\rm 
GP}(1,a_1)
\end{equation}
and the convergence is uniform on bounded intervals of $a_1$.
\end{thm}
\medskip

While we do not prove anything about Bose-Einstein condensation, which 
necessarily involves the full one-body density matrix $\rho^{(1)}(x,x')$, 
we can make an assertion about the diagonal part of the density 
matrix, $\rho^{\rm QM}(x)=\rho^{(1)}(x,x)$:

\begin{thm}[The GP density is the dilute limit of the QM density]\label{thm2}
For every fixed $a_1$
\begin{equation}
\lim_{N\to\infty}\frac{1}{N}\rho^{\rm QM}_{N,a_1/N}(x)=\rho_{1,a_1}^{\rm GP}(x)
\end{equation}
in the sense of weak convergence in $L^1$.
\end{thm}

For the proof of Theorem \ref{thm1} we establish upper and lower 
bounds on ${E^{\rm QM}(N,a)}$ in terms of $E^{\rm GP}(N,a)$ with 
controlled errors.  Theorem \ref{thm2} follows from Theorem \ref{thm1} 
by variation of the external potential.  The upper bound is obtained 
in Section 3 by a variational calculation which generalizes the upper 
bound of Dyson \cite{dyson} for a homogeneous gas of hard spheres.  We 
also derive an upper bound on the chemical potential, i.e., the 
energy increase when one particle is added to the system.  This upper 
bound is used in the proof of the lower bound of the energy in Section 
4.  The main ingredient for the lower bound, however, is the bound for 
the homogeneous case established in \cite{LY1998}.  In addition, some basic 
properties of the minimizer of the GP functional are used in the proof 
and we consider them next.

\section{The Gross-Pitaevskii Energy Functional}

The GP functional is defined by \eqref{gpfunct} for $\Phi\in\D$ with
\begin{equation}
\D=\{\Phi:\, \nabla\Phi\in L^2(\R^3), \V|\Phi|^2\in L^1(\R^3), 
\Phi\in L^4(\R^3)\cap L^2(\R^3)\},
\end{equation}
where $f\in L^p(\R^n)$ means $\int_{\R^n}|f(x)|^pd^nx<\infty$.
The corresponding GP energy is given by 
\begin{equation}\label{gpener}
E^{\rm GP}(N,a)=\inf\{\E^{\rm GP}[\Phi]:\, \Phi\in \D_N\}
\end{equation}
with
\begin{equation}
\D_N=\D\cap \{\Phi:\,
\hbox{$\int$}|\Phi(x)|^2d^3x=N\}.
\end{equation}

The basic facts about the GP functional are summarized in the following 
theorem.
\begin{thm}[Existence and properties of a minimizer]\label{GPtheo} 
The infi\-mum in \eqref{gpener} is a minimum, i.e., there is a $\Phi^{\rm 
GP}\in\D_N$ such that $E^{\rm GP}(N,a)=\E^{\rm GP}[\Phi^{\rm GP}]$. This 
$\Phi^{\rm GP}$ is unique up to a phase factor, which can be chosen so that 
$\Phi^{\rm GP}$ is strictly positive. $\Phi^{\rm GP}$ is at least once 
continuously differentiable, and if $\V$ is $C^\infty$ then also $\Phi^{\rm 
GP}$ 
is $C^\infty$. The energy $E^{\rm GP}(N,a)$ is continuously
differentiable in $a$ and hence (by Eq. (\ref{scaling})) also in $N$.
The minimizer $\Phi^{\rm GP}$ solves the Gross-Pitaevskii equation
\begin{equation}\label{gpglg}
-\nabla^2\Phi+\V\Phi+8\pi a|\Phi|^2\Phi=\mu \Phi
\end{equation}
(in the sense of distributions) with
\begin{equation}\label{mu}
\mu=dE^{\rm GP}(N,a)/dN=E^{\rm GP}(N,a)/N+4\pi a\bar \rho.
\end{equation}
Here $\bar\rho$  is the mean density \eqref{rhobar}.  
\end{thm}


The GP energy functional is mathematically  quite similar  to the energy 
functional of Thomas-Fermi-von Weizs\"acker theory 
and  Theorem \ref{GPtheo} can be proved by the methods 
of Sect.\ VII in \cite{lieb81}. For completeness, the proof is given in Appendix 
A. 
With additional properties of $\V$ one can draw further conclusions about 
$\Phi^{\rm GP}$:

\begin{prop}[Symmetry and monotonicity]
If $\V$ is spherically symmetric and monotone increasing, then $\Phi^{\rm GP}$ 
is 
spherically
symmetric and monotone decreasing.
\end{prop}
\begin{proof}
Let $\Phi^*$ be the symmetric-decreasing rearrangement of $\Phi^{\rm GP}$ (see
\cite{lieb}). Then $\E^{\rm GP}[\Phi^*]\leq \E^{\rm GP}[\Phi^{\rm GP}]$.
\end{proof}

\begin{prop}[Log concavity]
If $\V$ is convex, then $\Phi^{\rm GP}$ is log concave, i.e. 
$\Phi^{\rm GP}(x)^\lambda\Phi^{\rm GP}(y)^{(1-\lambda)}\leq 
\Phi^{\rm GP}\left(\lambda x+(1-\lambda)y\right)$, for all $x,y\in\R^3$, 
$\lambda\in (0,1)$.
\end{prop}
\begin{proof}
Using the Trotter product formula it suffices to show that the 
solutions 
$u(t,x)$ of the equations
\begin{displaymath}
\frac{\partial u}{\partial t}-\nabla^2 u=0, \quad
\frac{\partial u}{\partial t}+\V u=0,\quad
\frac{\partial u}{\partial t}+8\pi a u^3=\mu u 
\end{displaymath}
are log concave, if $u(0,x)$ is log concave. The first follows from the fact, 
that the convolution
of two log concave functions is log concave, the second follows easily from 
convexity of $\V$, and the third is
shown in \cite{lions}.
\end{proof}

The GP theory has a well defined limit if $Na\to\infty$. It is sometimes 
referred to as the \lq\lq Thomas-Fermi limit\rq\rq\ of GP theory because the 
gradient term vanishes in this limit. For simplicity we 
restrict ourselves to  {\it homogeneous external potentials} $\V$, i.e.,
\begin{equation}\label{homogen}\V(\lambda x)=\lambda^s\V(x)
\end{equation}
 for some $s>0$.

\begin{thm}[Large \textit{Na} limit]\label{TFlimit}
Let $V$ be homogeneous of order $s$ and let  $\FF$ be the functional
\begin{equation}
\FF[\rho]=\int_{\R^3}\left(\V(x)\rho(x)+4\pi a\rho(x)^2\right)d^3x
\end{equation}
with $\rho(x)\geq 0$, $x\in\R^3$. Let $F(N,a)$ be the infimum of $\FF$ under the 
condition 
$\int\rho=N$. By 
scaling, $F(N,a)=NF(1,Na)$ and $F(1,Na)=(Na)^{s/(s+3)}F(1,1)$. In the limit 
$Na\to\infty$ 
we have
\begin{equation}
\lim_{Na\to\infty}\frac{E^{\rm GP}(1,Na)}{(Na)^{s/(s+3)}}=F(1,1).
\end{equation}
The minimizing density of $\FF$ under the condition $\int\rho=1$ and
with $a=1$ 
is given by
\begin{equation}\rho^{\rm 
F}_{1,1}(x)=(8\pi)^{-1}\left[\widetilde\mu-\V(x)\right]_+
\end{equation}
with $\widetilde\mu=F(1,1)+4\pi\int\left(\rho^{\rm F}_{1,1}\right)^2$, 
and $[t]_{+}=t$ for $t> 0$ and $0$ otherwise. Moreover,
\begin{equation}
\lim_{Na\to\infty}\rho^{\rm GP}_{1,Na}(x)=\rho^{\rm F}_{1,1}(x)
\end{equation}
strongly in $L^2(\R^3)$.
\end{thm}
\begin{proof}
Since $\E^{\rm GP}[\sqrt\rho]\geq \FF[\rho]$ it is clear that 
$E^{\rm GP}(1,Na)\geq F(1,Na)$. For the converse we write
 $\rho$ in the form 
$\rho(x)=(Na)^{-3/(s+3)}\widetilde\rho\left((Na)^{-1/(s+3)}x\right)$
and obtain
\begin{align}\nonumber
\E^{\rm GP}[\sqrt{\rho}]&=(Na)^{s/(s+3)}\int\left((Na)^{-(s+2)/(s+3)}
\left|\nabla\sqrt{\widetilde\rho}\right|^2
+\V\widetilde\rho+4\pi\widetilde\rho^2\right)d^3x\\ \nonumber
\FF[\rho]&=(Na)^{s/(s+3)}\int\left(\V\widetilde\rho+
4\pi\widetilde\rho^2\right)d^3x.
\end{align}
In particular, $F(1,Na)=(Na)^{s/(s+3)}F(1,1)$, and with 
$\widetilde\rho=\rho^{\rm F}_{1,1}$ we obtain 
\begin{displaymath}
E^{\rm GP}(1,Na)\leq F(1,Na)+(Na)^{-2/(s+3)}\int\left|\nabla\sqrt{\rho^{\rm 
F}_{1,1}}\right|^2.
\end{displaymath}
(Regularizing $V$, if necessary, we may assume that 
$\int\left|\nabla\sqrt{\rho^{\rm 
F}_{1,1}}\right|^2<\infty$.) In the limit $Na\to\infty$ the gradient term 
vanishes, and thus the limit of the 
energies is proved. Now
\begin{displaymath}
\begin{split}
&\lim_{Na\to\infty}\frac{\E^{\rm GP}[\sqrt{\rho^{\rm 
GP}_{1,Na}}]}{(Na)^{s/(s+3)}}=\\
&\lim_{Na\to\infty}\left(\frac{\FF[\rho^{\rm GP}_{1,Na}]}{(Na)^{s/s+3)}}+
(Na)^{-\frac{s+2}{s+3}}\int\left|\nabla\sqrt{\rho^{\rm 
GP}_{1,Na}}\right|^2\right)=F(1,1).
\end{split}
\end{displaymath}
Since $F(1,1)$ is the minimum of $\FF/(Na)^{s/(s+3)}$, the second term
vanishes for $Na\to\infty$, and it follows that
$\rho^{\rm GP}_{1,Na}$ is a minimizing sequence for 
$\int\left(\V\rho+4\pi\rho^2\right)$. Since both terms in the functional are 
nonnegative, they must converge individually, in particular $\Vert \rho^{\rm 
GP}_{1,Na}\Vert_2$ converges to $\Vert \rho^{\rm 
F}_{1,1}\Vert_2$. On the other hand $\rho^{\rm GP}_{1,Na}$ converges weakly to 
$\rho^{\rm F}_{1,1}$ by uniqueness of the minimizer. Together with the 
convergence of the norms this implies strong convergence.

The solution of the variational equation for $\rho^{\rm F}_{1,1}$ is simply 
$\rho^{\rm F}_{1,1}=(8\pi)^{-1}[\widetilde\mu-\V]_+$ with $\widetilde\mu$ given 
by $\widetilde\mu=F(1,1)+4\pi\int\left(\rho^{\rm F}_{1,1}\right)^2$.
\end{proof}

\begin{lem}[Virial theorem]
When $V$ is homogeneous of order $s$, as in (\ref{homogen}), 
the minimizer of the GP functional satisfies
\begin{equation}\label{virial}
\frac{2}{3}\int|\nabla\Phi^{\rm GP}(x)|^2d^3x-\frac{s}{3}\int\Phi^{\rm 
GP}(x)^2\V(x)d^3x+4\pi a\int \Phi^{\rm GP}(x)^4d^3x=0.
\end{equation}
\end{lem}
\begin{proof}
Define $\Phi_k$ by
\begin{displaymath}
\Phi_k(x)=k^{1/2}\Phi^{\rm GP}(k^{1/3}x).
\end{displaymath}
Because $\Phi^{\rm GP}$ is the minimizer of $\E^{\rm GP}[\Phi]$, it must be 
true 
that
\begin{displaymath}
\left.\frac{\pd}{\pd k}\E^{\rm GP}[\Phi_k]\right|_{k=1}=0.
\end{displaymath}
This leads to the virial theorem \eqref{virial}.
\end{proof}

\medskip
In the proof of the lower bound we shall also consider the GP energy functional 
in a finite box. For $R>0$ we
denote by $\Lambda_R$ a cube centered at the origin, with side length $2R$. The 
energy functional 
$\E_R^{\rm GP}$ in the box is
simply \eqref{gpfunct} with the integration reduced to $\Lambda_R$, the 
corresponding minimizer, denoted by
$\Phi_R^{\rm GP}$, satisfies Neumann conditions at the boundary of
$\Lambda_R$, 
and is strictly positive.
Analogously to \eqref{gpdensity}, \eqref{rhobar} and \eqref{mu} 
we define $\rho_R^{\rm GP}$, $\bar\rho_R$ and $\mu_R$. The
corresponding energy will be denoted by $E_R^{\rm GP}(N,a)$. Then we have the 
following
\begin{lem}[GP energy in a box]\label{gpr}
\begin{equation}\label{limgpr}
\lim_{R\to\infty}E_R^{\rm GP}(N,a)=E^{\rm GP}(N,a)
\end{equation}\end{lem}
\begin{proof}
Using $N^{1/2}\Phi^{\rm GP}\chi_R/\|\Phi^{\rm GP}\chi_R\|_2$ as a test function 
for $\E_R^{\rm GP}$, 
where $\chi_R$ denotes the characteristic function of $\Lambda_R$, we 
immediately  get
\begin{equation}\label{bound}
\lim_{R\to\infty}E_R^{\rm GP}(N,a)\leq E^{\rm GP}(N,a).
\end{equation}
Let $\Theta_R$ be a $C^\infty$ function on $\R^3$, with $\Theta_R=0$ outside 
$\Lambda_R$, and 
$\Theta_R=1$ inside $\Lambda_{R-1}$. We use 
$N^{1/2}\Phi_R^{\rm GP}\Theta_R/\|\Phi_R^{\rm GP}\Theta_R\|_2$ as a test 
function for 
$\E^{\rm GP}$. Since $\nabla\Theta_R$ is bounded and
\begin{displaymath}
\lim_{R\to\infty}\inli_{\Lambda_R\setminus\Lambda_{R-1}}|\Phi_R^{\rm GP}|^2=0
\end{displaymath}
(because $\V$ tends to infinity and $\E^{\rm GP}_R[\Phi^{\rm GP}_R]$ is bounded 
by \eqref{bound}), we have
\begin{displaymath}
E^{\rm GP}(N,a)\leq\liminf_{R\to\infty}E_R^{\rm GP}(N,a).
\end{displaymath}
This completes the proof of \eqref{limgpr}.
\end{proof}

\section{Upper Bounds}
\subsection{Upper bound for the QM energy}
It will now be shown that for all $N$ and small values of
$a\bar\rho^{1/3}$ (with $\bar\rho=\int\rho^{\rm GP}(x)^2d^3x/N$, cf.\ 
\eqref{rhobar})
\begin{equation}\label{upper}
E^{\rm QM}(N,a)\leq
E^{\rm GP}(N,a)\left(1+O\left(a\bar\rho^{\frac{1}{3}}\right)\right).
\end{equation}
This upper bound, which holds for all positive, spherically symmetric $v$ with 
finite scattering length, is derived by means of the variational principle. We 
 generalize
a method of Dyson \cite{dyson}, who proved an upper bound for the homogeneous
Bose gas with
hard-sphere interaction. Consider as a trial function
\begin{gather}
\Psi=\F(x_1,\dots, x_N)\G(x_1,\dots, x_N)\\ \intertext{with} \F(x_1,\dots,
x_N)=\mbox{$\prod_{i=1}^N$} F_i(x_1,\dots, x_i), \quad \G(x_1,\dots, x_N)= 
\mbox{$\prod_{i=1}^N$} 
g(x_i)\\ 
\intertext{where} F_i(x_1,\dots,x_i)=f(t_i), 
\quad t_i=\min\left(\xij,j=1,\dots, 
i-1\right),\label{deft}\\ \intertext{with a function $f$ satisfying}  0\leq 
f\leq 1,
\quad f'\geq 0,\label{propf} \\ \intertext{and} g(x)=\Phi^{\rm 
GP}(x)/\|\Phi^{\rm 
GP}\|_\infty.
\end{gather}
The function $f$ will be specified later.
This trial function is not symmetric in the particle coordinates, but the
expectation value $\langle\Psi|H^{(N)}\Psi\rangle /\langle\Psi|\Psi\rangle$ is 
still an
upper bound to the bosonic ground state energy because the Hamiltonian is 
symmetric and its ground state wave function is positive. Hence the
bosonic ground state energy is equal to the {\it absolute} ground
state energy (\cite{dyson}, \cite{lieb63}).

The physical meaning of the trial function can be understood as follows:  The 
$G$ part describes independent particles, each with the GP wave function. The 
$F$ part means that the particles are inserted into the system one at a time, 
taking into account the particles previously inserted, but without adjusting 
their wave function (cf.\ \cite{dyson}). Although a wave function of this form 
cannot describe all correlations present in the true ground state, it captures 
the leading term in the energy for dilute systems.
 
For the computation of the kinetic energy we
use
\begin{equation}\label{laplace}
\int_{\R^{3N}}\Psi\nabla_k^2\Psi=\int_{\R^{3N}}\left(\G\nabla_k^2\G\right)\F^2-
\int_{\R^{3N}}\G^2|\nabla_k\F|^2,
\end{equation}
where $\nabla_k$ denotes the gradient with respect to $x_k$, $k=1,\dots,N$.  We write
\begin{equation}
\eps_{ik} =\begin{cases}
1 &\text{for $i=k$}\\ 
-1 &\text{for $t_i=|x_i-x_k|$}\\ 
0 &\text{otherwise}
\end{cases}.
\end{equation}
Let $n_i$ be the unit vector in the direction of $x_i-x_{j(i)}$, when 
$x_{j(i)}$ 
is
the nearest to $x_i$ of the points $(x_1,\dots, x_{i-1})$. (Note that ${j(i)}$ really
depends on all the points $x_1,\dots,x_i$ and not just on the index 
$i$. Except for a set of zero measure, $j(i)$ is unique.) Then
\begin{equation}
\G\nabla_k\F=\suli_i\Psi F_i^{-1}\eps_{ik}n_i f'(t_i),
\end{equation}
and after summation over $k$
\begin{equation}\label{term}
\begin{split}
&\suli_k\G^2|\nabla_k\F|^2=|\Psi|^2\suli_{i,j,k}\eps_{ik}\eps_{jk}(n_i\cdot 
n_j)
F_i^{-1}F_j^{-1}f'(t_i)f'(t_j)\\ &\leq 2|\Psi|^2\suli_i 
F_i^{-2}f'(t_i)^2+2|\Psi|^2
\suli_{k\leq i<j}|\eps_{ik}\eps_{jk}|F_i^{-1}F_j^{-1}f'(t_i)f'(t_j).
\end{split}
\end{equation} The
expectation value can thus be bounded as follows:
\begin{equation}\label{exp}
\begin{split}
\frac{\langle\Psi|H^{(N)}\Psi\rangle}{\langle\Psi|\Psi\rangle}&\leq
2\suli_{i=1}^N\frac{\int |\Psi|^2F_i^{-2}f'(t_i)^2}{\int
|\Psi|^2}+\suli_{j<i}\frac{\int |\Psi|^2v(\xij)}{ \int |\Psi|^2}\\ 
&+2\suli_{k\leq
i<j}\frac{\int 
|\Psi|^2|\eps_{ik}\eps_{jk}|F_i^{-1}F_j^{-1}f'(t_i)f'(t_j)}{\int
|\Psi|^2}\\ &+\suli_{i=1}^N\frac{\int |\Psi|^2\left(-g(x_i)^{-1}\nabla^2_i
g(x_i)+\V(x_i)\right)}{\int |\Psi|^2}.
\end{split}
\end{equation}
For $i<p$, let $F_{p,i}$ be the value that $F_p$ would take 
if the point $x_i $ were
omitted, i.e.,
\begin{equation}
F_{p,i}=f(|x_p-x_{k(p)}|),
\end{equation}
where $x_{k(p)}$ is the nearest to $x_p$ of the points $(x_1,\dots,
x_{i-1},x_{i+1},\dots, x_{p-1})$. The reason for introducing these functions is 
that one wants to decouple the integration over $x_i$ from the integrations 
over the other variables. (Note that $F_{p,i}$ is independent of $x_i$.) 
Analogously, one defines $F_{p,ij}$ by omitting $x_i$ and $x_j$. This 
decouples simultaneously $x_i$ and $x_j$ from the other variables. The functions 
$F_i$ occur both in the numerator and the denominator so one needs estimates 
from below and above. Since
\begin{equation}
F_p=\min\{F_{p,ij},f(|x_p-x_j|),f(|x_p-x_i|)\},
\end{equation}
one has (recall that $f\leq 1$)
\begin{equation}
F_{p,ij}^2f(|x_p-x_i|)^2f(|x_p-x_j|)^2\leq F_p^2\leq F_{p,ij}^2.
\end{equation}
Hence, with $j<i$,
\begin{align} 
&F_{j+1}^2\dots F_{i-1}^2F_{i+1}^2\dots F_N^2\leq F_{j+1,j}^2\dots
F_{i-1,j}^2F_{i+1,ij}^2\dots F_{N,ij}^2 \label{above}\\ 
\intertext{and}\label{below} &F_j^2\dots F_N^2\geq
F_{j+1,j}^2\dots F_{i-1,j}^2F_{i+1,ij}^2\dots F_{N,ij}^2\\ \nonumber
&\cdot\left(1-\suli_{k=1,\, k\neq i,j}^N(1- f(|x_j-x_k|)^2) 
\right)
\cdot\left(1-\suli_{k=1,\, k\neq i}^N(1-f(|x_i-x_k|)^2)\right).
\end{align}
We now consider the first two terms in \eqref{exp}. 
In the numerator of the first term for each fixed $i$ we use the estimate
\begin{equation}
f'(t_i)^2\leq\suli_{j=1}^{i-1} f'(\xij)^2,
\end{equation}
and in the second term we use $F_i\leq f(|x_i-x_j|)$. For fixed $i$ and $j$ 
one
eliminates $x_i$ and 
$x_j$
from the rest of the integrand by using \eqref{above} and $F_j\leq 1$ in the 
numerator and \eqref{below} in the denominator to do the 
$x_i$
and $x_j$ integrations. With the transformation $\eta=x_i-x_j$,
$\chi=(x_i+x_j)/2$ one gets
\begin{equation}\label{trafo}
\begin{split}
&\int \left(2f'(\xij)^2+v(\xij)f(\xij)^2\right)g(x_i)^2g(x_j)^2d^3x_i d^3x_j=\\ 
&\int
\left(2f'(|\eta|)^2+v(|\eta|)f(|\eta|)^2\right)g(\chi+\half \eta)^2
g(\chi-\half
\eta)^2 d^3\eta d^3\chi.
\end{split}
\end{equation}
By the Cauchy-Schwarz inequality 
\begin{equation}
\int g(\chi+\half \eta)^2g(\chi-\half \eta)^2 d^3\chi\leq \int g(\chi)^4
d^3\chi,
\end{equation}
and one obtains
\begin{equation}\label{j}
\begin{split}
&\int \left(2f'(\xij)^2+v(\xij)f(\xij)^2\right)g(x_i)^2g(x_j)^2d^3x_i d^3x_j\\ 
&\leq \int
g(\chi)^4 d^3\chi \int \left(2f'(|\eta|)^2+v(|\eta|)f(|\eta|)^2\right) 
d^3\eta\equiv
2\int g(\chi)^4 d^3\chi J.
\end{split}
\end{equation}
In the denominator one gets, using that $0\leq g\leq 1$,
\begin{equation}\label{i}
\begin{split}
&\int\left(1-\suli_{p=1,\,p\neq i}^N(1-f(|x_p-x_i|)^2)\right)g(x_i)^2d^3x_i\\ 
&\geq
\int
g(x)^2 d^3x-N\int(1-f(|x_p-x_i|)^2) \equiv \int g(x)^2 d^3x-NI.
\end{split}
\end{equation}
The same factor comes from the $x_j$-integration, the remaining factors are 
identical
in numerator and denominator, and so finally the first and second term are 
bounded by
\begin{equation}\label{bound1}
\suli_{i=1}^N(i-1) \frac{2\int g(x)^4 d^3x J}{(\int g(x)^2 d^3x-NI)^2}\leq N^2 
\frac{\int
g(x)^4 d^3x J}{(\int g(x)^2 d^3x-NI)^2}.
\end{equation}

A similar argument is now applied to the third term of \eqref{exp}. Note that 
the
contributions from $k=i$ and $k<i$ are the same. Therefore
\begin{equation}
\begin{split}
&\suli_{k=1}^i\int |\eps_{ik}|f(t_i)f'(t_i)|\eps_{jk}|f(t_j)f'(t_j)g(x_i
)^2g(x_j)^2d^3x_id^3x_j\leq \\ &2\suli_{k=i}^{i-1}\int f(|x_i-x_k|)f'(|x_i-x_k|
)f(|x_j-x_k|)f'(|x_j-x_k|)g(x_i)^2g(x_j)^2d^3x_id^3x_j.
\end{split}
\end{equation}
With $g\leq 1$ one gets
\begin{equation}\label{k}
\begin{split}
&2\suli_{k=1}^{i-1}\int f(|x_i|)f'(|x_i|)f(|x_j|)f'(|x_j|)d^3x_i d^3x_j\\
&=2(i-1)\left(\int f(|x|)f'(|x|)d^3x\right)^2\equiv 2(i-1)K^2.
\end{split}
\end{equation}
The summation over $i$ and $j$ gives
\begin{equation}
\suli_{j=1}^N \suli_{i=1}^{j-1} (i-1)=\frac{1}{6}N(N-1)(N-2).
\end{equation}
In the denominator one gets the same factors as above, and so the third 
term
is bounded by
\begin{equation}
\frac{2}{3}N^3\frac{K^2}{(\int g(x)^2 d^3x-NI)^2}.
\end{equation}

Next consider the last term of \eqref{exp}. Define $\e$ by
\begin{equation}\label{e}
\int\left(-g(x)\nabla^2 g(x)+\V(x)g(x)^2\right)d^3x\equiv\e\int g(x)^2d^3x.
\end{equation}
After eliminating $x_i$ from the integrands in the numerator and the
denominator and
using $F_i\leq 1$ one sees that the term is bounded above by
\begin{equation}
N\frac{\e\int g(x)^2 d^3x}{\int g(x)^2 d^3x-NI}.
\end{equation}

Putting all terms together we obtain as an upper bound for the ground state
energy
\begin{equation}\label{ener}
\frac{E^{\rm QM}}{N}\leq \frac{\e\int g(x)^2 d^3x}{\int g(x)^2 d^3x-NI}+N 
\frac{\int g(x)^4 d^3x 
J}{(\int
g(x)^2 d^3x-NI)^2}+ \frac{2}{3}N^2\frac{K^2}{(\int g(x)^2 d^3x-NI)^2}
\end{equation}
with $I$, $J$, $K$ and $\e$ defined by \eqref{i}, \eqref{j}, \eqref{k} and 
\eqref{e}. It remains to choose $f$. We take for $b>a$ (we shall soon fix $b$)
\begin{equation}\label{deff}
f(r)=\begin{cases} (1+\eps)u(r)/r &\text{for $r\leq b$}\\ 1
&\text{for
$r>b$}
\end{cases}
\end{equation}
where $u(r)$ is the solution of the scattering equation
\begin{equation}
-u''(r)+\half v(r)u(r)=0 \quad\text{ with $u(0)=0$,
$\lim_{r\to\infty}u'(r)=1$}
\end{equation}
and $\eps$ is determined by requiring $f$ to be continuous. Convexity of $u$ 
gives
\begin{equation}
r\geq u(r)\geq \begin{cases}
0& \text{for $r\leq a$}\\
r-a& \text{for $r>a$}
\end{cases}
,\quad 1\geq u'(r) \geq \begin{cases} 0& \text{for $r\leq a$}\\ 
1-\frac{a}{r}&
\text{for $r>a$}
\end{cases}.
\end{equation}
These estimates imply
\begin{align}\label{imply}
&J\leq (1+\eps)^2 4\pi a\\ &I\leq 4\pi\left(\frac{a^3}{3}+ab(b-a)\right)\\ 
&K\leq 4\pi
(1+\eps)a\left(b-\frac{a}{2}\right)\\ 
&0\leq\eps\leq\frac{a}{b-a}.\label{imply2}
\end{align}
For \eqref{imply} we used partial integration. By definition 
\eqref{rhobar} 
\begin{equation}
\bar\rho=\frac{1}{N}\int(\rho^{\rm GP})^2=N\frac{\int g^4}{\left(\int
g^2\right)^2},
\end{equation}
and we choose $b$ such that
\begin{equation}\label{defc}
\frac{4\pi}{3}\bar\rho b^3=\frac{\int g^4}{\int
g^2}=\frac{\bar\rho}{\|\rho^{\rm GP}\|_\infty}\equiv c.
\end{equation}
With this choice the factor in the denominators in \eqref{ener} is bounded by
\begin{equation}\label{gi}
\int g^2 -NI\geq\int g^2\left(1-\frac {a} {b}\right)^3.
\end{equation}
Note that $c\leq 1$,  and $a<b$ holds provided
\begin{equation}
\frac{a^3}{b^3}=\frac{4\pi}{3}a^3{\|\rho^{\rm GP}\|_\infty}<1.
\end{equation}
Collecting the estimates \eqref{imply}-\eqref{imply2},  we finally obtain

\begin{thm}[Upper bound for the QM energy]\label{upperthm}
\begin{equation}\label{erg} 
{E^{\rm QM}}\leq\frac{\int \left(|\nabla \Phi^{\rm GP}|^2+\V(\Phi^{\rm 
GP})^2\right)}{(1-\frac{a}{b})^3}+ 4\pi a\int (\Phi^{\rm GP})^4
\frac{1+\frac{2}{c}\aob-\frac{2}{c}\aob^2+\frac{1}{2c}\aob^3}{
(1-\frac{a}{b})^8}
\end{equation}
with $b$ and $c$ defined by \eqref{defc}.
\end{thm}

\begin{rem}[Negative potentials with hard core]
\eqref{upper} can be extended to include partially negative potentials of the
form
\begin{equation}\label{negpot}
v(r)=\begin{cases}
\infty &\text{for $0\leq r\leq d$}\\
-|w(r)| &\text{for $d<r\leq R_0$}\\
0 &\text{for $r>R_0$},
\end{cases}
\end{equation}
as long as $f^\prime(r)^2+\frac{1}{2}v(r)f(r)^2\geq 0$ for all $r$. With $f$ 
from
\eqref{deff}, this is the case for sufficient shallow potentials. 
The potential 
energy
is then negative, and the estimates used for \eqref{trafo} are no longer valid. But
\begin{equation}
\suli_{j<i} v(|x_i-x_j|)F_i^2 \leq\suli_i v(t_i)F_i^2,
\end{equation}
and because of $2f^\prime(r)^2+v(r)f(r)^2 \geq 0$ we get
\begin{equation}
2f^\prime(t_i)^2+v(t_i)f(t_i)^2 
\leq\suli_{j=1}^{i-1}2f^\prime(|x_i-x_j|)^2+
v(|x_i-x_j|)f(|x_i-x_j|)^2.
\end{equation}
So we have the same as in \eqref{trafo}. Note that for potentials as in 
\eqref{negpot} $f$ satisfies \eqref{propf}, as long as $a>0$.
\end{rem}

\begin{rem}[Homogeneous gas]
For the special case of a homogeneous Bose gas (i.e. $\V=0$) in a box
of 
volume ${\cal 
V}$, the
GP density is simply
\begin{equation}
\rho^{\rm GP}(x)=N/{\cal V}=\bar\rho,
\end{equation}
for all $x$ in the box, and the GP energy is given by
\begin{equation}
E^{\rm GP}(N,a)=4\pi a\frac{N^2}{{\cal V}}.
\end{equation}
Our method also applies here, if we impose periodic boundary conditions on
the box.
Therefore our upper bound is a generalization of a result by Dyson
\cite{dyson}, who proved an
analogous bound for the special case of a homogeneous Bose gas with hard-sphere
interaction.
\end{rem}

\subsection{Upper bound for the chemical potential}
By the same method as in the previous subsection one can derive a bound on the
increase of the energy when one particle is added to the system. This bound 
will be needed for the derivation of the lower bound to the energy.

\begin{thm}[Upper bound for the chemical potential]\label{chempot}
Let $E^*(N,a)$ denote the infimum of the functional
\begin{equation}
\E^*[\Phi]=\int\left(|\nabla\Phi(x)|^2+\V(x)|\Phi(x)|^2+8\pi
a\|\Phi\|_\infty^2|\Phi(x)|^2\right)d^3x
\end{equation}
with $\int |\Phi|^2=N$. Let $\Phi^*$ be the positive minimizer of $\E^*$ (its 
existence is 
guaranteed by the same arguments as for the GP functional itself), and 
$\bar\rho^*=\int\Phi^{*4}/N$.
Then
\begin{equation}\label{n+1-n}
E^{\rm QM}(N+1,a)\leq E^{\rm 
QM}(N,a)+E^*(1,Na)\left(1+O(a\bar\rho^{*1/3})\right).
\end{equation}
\end{thm}
\begin{proof}
Let $\Psi^{(N)}_0$ be the ground state  wave function of $H^{(N)}$. As test 
wave function for $H^{(N+1)}$ we take
\begin{equation}\label{Test}
\Psi(x_1,\dots, x_{N+1})=\Psi^{(N)}_0(x_1,\dots, x_N)\Phi^*(x_{N+1}) 
f(t_{N+1}),
\end{equation}
where $f$ and $t_{N+1}$ are defined as in \eqref{deff} and \eqref{deft}, i.e., 
$rf(r)$ is essentially the zero energy scattering solution and $t_{N+1}$ 
is the distance of $x_{N+1}$ from its nearest neighbor. 
We have
\begin{equation}\label{langle}
\begin{split}
&\langle\Psi|H^{(N+1)}\Psi\rangle=\\ &\int
f^2\Phi^{*2}\left(\suli_{i=1}^N\left(-\overline{\Psi^{(N)}_0}\nabla^2_i\Psi^{(N)
}_0
+\V(x_i)|\Psi^{(N)}_0|^2\right)
+\suli_{i<j}^N v(|x_i-x_j|)|\Psi^{(N)}_0|^2\right)\\ &+\int
|\Psi^{(N)}_0|^2\Phi^{*2}\left(\suli_{i=1}^N|\nabla_i f|^2\right) 
+\int|\Psi^{(N)}_0|^2
f^2\left(-\Phi^*\nabla^2_{N+1}\Phi^*+\V(x_{N+1})\Phi^{*2}\right)\\
&+\int|\Psi^{(N)}_0|^2\Phi^{*2}\left(|\nabla_{N+1}f|^2 +\suli_{i=1}^N
v(|x_{N+1}-x_i|)f^2\right).
\end{split}
\end{equation}
For $f$ one uses the estimates
\begin{equation}
f\leq 1,\quad f(t_{N+1})^2\geq 
1-\suli_{i=1}^N\left(1-f(|x_{N+1}-x_i|)^2\right),
\end{equation}
and for the derivatives one has
\begin{gather}
|\nabla_{N+1}f|^2=f'(t_{N+1})^2=\suli_{i=1}^N |\nabla_i f|^2,\\ 
\intertext{and}
f'(t_{N+1})^2\leq\suli_{i=1}^N f'(|x_{N+1}-x_i|)^2.
\end{gather}
After division by the norm of $\Psi$ \eqref{langle} becomes
\begin{equation}
\begin{split}
&E^{\rm QM}(N+1,a)\leq
E^{\rm 
QM}(N,a)+\frac{\int|\Psi^{(N)}_0|^2\left(-\Phi^*\nabla^2_{N+1}\Phi^*+\V\Phi^{*2}
\right)}
{\int|\Psi^{(N)}_0|^2\left(N-N\Phi^{*2}\int(1-f^2)\right)}\\
&+\frac{\int|\Psi^{(N)}_0|^2\|\Phi^*\|_\infty^2\suli_{i=1}^N\left(2f'(|x_{N+1}
-x_i|)^2+v(|x_{N+1}-x_i|)f(|x_{N+1}-x_i|)^2\right)}
{\int|\Psi^{(N)}_0|^2\left(N-N\|\Phi^*\|_\infty^2\int(1-f^2)\right)}.
\end{split}
\end{equation}
$\Psi^{(N)}_0$ does not depend on $x_{N+1}$. One integrates first over
$x_{N+1}$ and then over the remaining variables. In analogy with the estimates 
\eqref{imply} 
and \eqref{gi} one gets
\begin{gather}
\int(2f'(|x|)^2+v(|x|)f(|x|)^2)d^3x\leq 8\pi 
a\left(1+O(a\bar\rho^{*1/3})\right),\\ 
\intertext{and}\|\Phi^*\|_\infty^2\int(1-f(|x|)^2)d^3x\leq 
O(a\bar\rho^{*1/3}).
\end{gather}
This implies
\begin{equation}
E^{\rm QM}(N+1,a)\leq E^{\rm 
QM}(N,a)+\frac{E^*(N,a)}{N}\left(1+O(a\bar\rho^{*1/3})\right).
\end{equation}
By scaling,  $E^*(N,a)=N E^*(1,Na)$ and \eqref{n+1-n} follows.
\end{proof}

\medskip
\begin{rem}[Box with Neumann conditions]
Eq.\ \eqref{n+1-n} also holds for a box with Neumann boundary conditions. To see this 
we note that 
Neumann conditions give the lowest energy for the quadratic form 
$\langle\Psi|H^{(N+1)}\Psi\rangle$, therefore it is
possible to use \eqref{Test} as a test function, even if $f$ does not fulfill 
Neumann conditions. If
$\Psi_0^{(N)}$ and $\Phi^*$ do, the calculation above is still valid, since for
\begin{equation}
\int|\nabla (gf)|^2=-\int f^2 g\nabla^2 g+\int g^2 |\nabla f|^2
\end{equation}
only boundary conditions for $g$ are needed.

Note also that in the homogeneous case, i.e. $\V=0$ in the box, 
$E^*(N,a)=2E^{\rm GP}(N,a)$.
\end{rem}

\section{Lower Bounds}

\subsection{The homogeneous case}

In \cite{LY1998} the following lower bound was established for 
the ground state energy, $E^{\rm hom}$, of a Bose gas
in a box of side length $L$ with Neumann boundary conditions and $v$ of finite 
range:
\begin{equation}\label{basic}
E^{\rm hom}(N,L)\geq 4\pi a\frac {N^2} {L^3}(1-CY^{1/17})
\end{equation}
with $Y=a^3N/L^3$ and $C$ a constant. The 
estimate holds for all $Y$ small enough and $L/a\gg Y^{-6/17}$ (note that
this implies $N\gg Y^{-1/17}$). In the thermodynamic limit the constant $C$ can 
be taken to be $C=8.9$, but this value is only of academic interest, because 
the error term $-CY^{1/17}$ is not believed to reflect the true state of 
affairs. Presumably, it does not even have the correct sign.

The restriction of a finite range can be relaxed. In fact, 
 \eqref{basic} holds (with a different constant $C$) for all positive, 
spherically symmetric $v$ with
\begin{equation}\label{condition}
v(r)\leq {\rm const.\,}r^{-(3+\frac 15+\epsilon)}\qquad\text{ for $r$ large 
enough, $\epsilon>0$.}
\end{equation}
More generally, if  
 \begin{equation}\label{condition2}
v(r)\leq {\rm const.\,}r^{-(3+\epsilon)}\qquad\text{ for $r$ large enough, 
$\epsilon>0$}
\end{equation}
then \eqref{basic} holds at least with the exponent $1/17$ replaced by an 
exponent
$O(\epsilon)$. We prove these assertions in Appendix B.

In the next section we shall stick to the estimate 
\eqref{basic} for simplicity, but in the limit
$N\to\infty$ the explicit form of the error 
term is not significant so a decrease of the potential as in 
\eqref{condition2} is sufficient for the limit Theorems \ref{thm1} and 
\ref{thm2}.

\subsection{The lower bound in the inhomogeneous case}

Our generalization of \eqref{basic} to the inhomogeneous case is as follows:

\begin{thm}[Lower bound for the QM energy]\label{lowerthm}
Let $v$ be positive, spherically symmetric and decrease at infinity like  
\eqref{condition}. Its scattering length is $a=a_1/N$ with $a_1$ fixed, as 
explained in the Introduction. Then as $N\to\infty$
\begin{equation}\label{lower}
E^{\rm QM}(N,a)\geq E_R^{\rm GP}(N,a)\left(1-\emph{const.\,}N^{-1/10}\right)
\end{equation}
for all $R$ large enough, where $E_R^{\rm GP}$ is the GP energy in a cube with 
side length 
$2R$, center at the origin, and Neumann boundary conditions; the constant in 
\eqref{lower}
depends only on $a_1$ and $R$.

\end{thm}

\begin{proof}
As in \cite{LY1998} the lower bound will be obtained by 
dividing space into cubic boxes with Neumann conditions at the 
boundary, which only lowers the energy.  Moreover, interactions among 
particles in different boxes are dropped.  Since $v\geq 0$, this, too,
lowers the energy.  For the lower bound one has to estimate the 
energy for a definite particle number in each box and then to optimize 
over all distributions of the $N$ particles among the boxes.  

\medskip
\textit{Step 1 (Finite box):} The first step is to show that all the particles 
can be 
assumed to be in some large, but finite
box. Since
\begin{equation}
K(R)=\inf_{|x|>R}\V(x)
\end{equation}
tends monotonically to $\infty$ with $R$, one knows that the energy of a 
particle outside a cube $\Lambda_R$ of side length $2R$ and center at the origin 
 is 
at least $K(R)$. Hence
\begin{equation}
E^{\rm QM}(N,a)\geq \inf_{0\leq n\leq N}\{E^{\rm QM}_R(N-n,a)+nK(R)\},
\end{equation}
where $E^{\rm QM}_R(N-n,a)$ denotes the energy of $N-n$ particles in 
$\Lambda_R$, with Neumann conditions at
the boundary.
We now apply Theorem \ref{chempot} (which holds also in a cube with Neumann 
conditions). Applying the theorem $n$ times and noting that $E^*(1,Na)$ is 
monotone in $N$ we have
\begin{equation}
E^{\rm QM}_R(N-n,a)\geq E^{\rm 
QM}_R(N,a)-nE^*(1,Na)\left(1+O(a\bar\rho^{*1/3})\right).
\end{equation}
Hence there is a constant $K'$ (that depends only on $Na$), such that
$K(R)>K'$
implies that the infimum is obtained at $n=0$. This is fulfilled for all 
sufficiently large  $R$, independently of $N$ if $Na$ is fixed. So we can 
restrict ourselves 
to estimating the energy in $\Lambda_R$ with Neumann boundary conditions.

\medskip \textit{Step 2 (Trading $\V$ for $-\rho_R^{\rm GP}$):} We 
shall now use the GP equation to eliminate $\V$ from the problem, 
effectively replacing it by $-8\pi a\rho_R^{\rm GP}$.   We write the 
wave function in $\Lambda_R^N$ as
\begin{equation}
\Psi(x_1,\dots,x_N)=f(x_1,\dots, x_{N})\prod_{i=1}^{N}\Phi_R^{\rm GP}(x_i),
\end{equation}
where $\Phi_R^{\rm GP}$ denotes the the minimizer of the GP functional in 
$\Lambda_R$; since it 
is strictly positive, every wave function can be written in this form. 
Note also that $\Phi_R^{\rm GP}$ and $f$ obey Neumann conditions. We have
\begin{equation}
\begin{split}
\langle\Psi|H\Psi\rangle&=\suli_{i=1}^{N}\int|\Psi|^2\Phi_R
^{\rm GP}(x_i)^{-1}
\left(-\nabla^2_i+\V(x_i)\right)\Phi_R^{\rm GP}(x_i)d^{3N}x\\
&+\suli_{i=1}^{N}\int\prod_{k=1}^{N}\Phi_R^{\rm GP}(x_k)^2|\nabla_i
f|^2d^{3N}x+\suli_{i<j}^{N}\int|\Psi|^2v(\xij)d^{3N}x,
\end{split}
\end{equation}
where the integrals are over $\Lambda_R^N$. Using the GP equation \eqref{gpglg}
this becomes
\begin{equation}
\begin{split}
&\langle\Psi|H\Psi\rangle=\\
&\suli_{i=1}^{N}\int\prod_{k=1}^{N}\rho_R^{\rm GP}(x_k) \left(|\nabla_i 
f|^2+
\left(\mu_R-8\pi a\rho_R^{\rm GP}(x_i)+\sum_{j=1}^{i-1} 
v(|x_i-x_j|)\right)|f|^2\right).
\end{split}
\end{equation}
Inserting the value \eqref{mu} for $\mu_R$ gives
\begin{equation}\label{ener2}
\frac{\langle\Psi|H\Psi\rangle}{\langle\Psi|\Psi\rangle}-E_R^{\rm GP}=4\pi a
\bar\rho_R N+Q(f)
\end{equation}
\begin{equation}\label{ener3}
Q(f)=\suli_{i=1}^{N}
\frac{\inli_{\Lambda_R^N}\prod\limits_{k=1}^{N}\rho_R^{\rm GP}(x_k) 
\left(|\nabla_i 
f|^2+\suli_{j=1}^{i-1}
v(|x_i-x_j|)|f|^2-8\pi a\rho_R^{\rm GP}(x_i)|f|^2\right)}
{\inli_{\Lambda_R^N}\prod\limits_{k=1}^{N}\rho_R
^{\rm GP}(x_k)|f|^2}.
\end{equation}

\medskip
\textit{Step 3 (Division into boxes):}
$Q(f)$ is a normalized quadratic form on the weighted $L^2$-space
$L^2(\Lambda_R^N,\prod_{k=1}^N\rho_R^{\rm GP}(x_k)d^3x_k)$, and can be 
minimized by dividing the cube $\Lambda_R$ into smaller cubes 
with side length L, labelled by $\al$, distributing the $N$ particles among the 
boxes and 
optimizing over all distributions. We therefore have
\begin{equation}
\inf_f Q(f)\geq \inf_{\{n_\al\}} \suli_\al \inf_{f_\al}Q_\al (f_\al),
\end{equation}
where the infimum is taken over all distributions of the particles with 
\mbox{$\sum n_\al=N$},
and $Q_\al (f)$ is given by
\begin{equation}\label{qal}
Q_\al (f)=\suli_{i=1}^{n_\al}
\frac{\inli_\al\prod\limits_{k=1}^{n_\al}\rho_R^{\rm GP}(x_k) \left(|\nabla_i 
f|^2+\suli_{j=1}^{i-1}
v(|x_i-x_j|)|f|^2-8\pi a\rho_R^{\rm GP}(x_i)|f|^2\right)}{\inli_\al
\prod\limits_{k=1}^{n_\al}\rho_R
^{\rm GP}(x_k)|f|^2},
\end{equation}
where the integrals are over $x_k$ in the box $\al$,
$k=1,\dots,n_\alpha$. 
Note that here $f=f(x_1,\dots, 
x_{n_\al})$, and \eqref{qal} is the same as \eqref{ener3} with $N$ replaced by 
$n_\al$ and $\Lambda_R$ with the box $\al$.

We now want to use \eqref{basic} and therefore we must approximate  
$\rho_R^{\rm GP}$ by constants in each box. Let 
$\rmax$ and
$\rmin$, respectively, denote the maximal and minimal values of $\rho_R^{\rm GP}$ 
in box $\al$. 
With
\begin{equation}
\Phi^{(i)}(x_1,\dots, x_{n_\al})=f(x_1,\dots, x_{n_\al}) 
\prod_{\substack{k=1 \\ k\neq 
i}}^{n_\al}\Phi_R^{\rm GP}(x_k),
\end{equation}
one has
\begin{equation}
\begin{split}
&\frac{\int\prod_k\rho_R^{\rm GP}(x_k)\left(|\nabla_i 
f|^2+\sum_{j=1}^{i-1}
v(|x_i-x_j|)|f|^2\right)} {\int\prod_k \rho_R^{\rm GP}(x_k)|f|^2}
\\&\geq
\frac{\rmin}{\rmax}\frac{\int |\nabla_i\Phi^{(i)}|^2 +\sum_{j=1}^{i-1}
v(|x_i-x_j|)|\Phi^{(i)}|^2}{\int|\Phi^{(i)}|^2}.
\end{split}
\end{equation}
This holds for all $i$, and if we use 
$\rho_R^{\rm GP}(x_i)\leq \rmax$ in \eqref{qal}, we get
\begin{equation}
Q_\al(f)\geq \frac{\rmin}{\rmax}E^{\rm hom}(n_\al,L)-8\pi a\rmax n_\al,
\end{equation}
where $E^{\rm hom}$ is the energy in a box without an external potential.

{\it Remark:\/} If we had not taken Step 2 and used instead the division into 
boxes directly on  the original Hamiltonian \eqref{ham} we would 
be considering  the minimization of
\begin{equation}\label{wrongmin}
\sum_{\alpha}
E^{\rm hom}(n_\al,L)+V_{\alpha,{\rm min}}n_{\alpha}.    
\end{equation}
Such a procedure, however, would not lead to the GP energy. To see this, 
consider the special case of no interaction, i.e.,  
$v=0$ and hence also also $E^{\rm 
hom}(n_\al,L)=0$. The minimum of \eqref{wrongmin} is then simply 
$N\min_{x} V(x)$, whereas the GP energy is in this case $N$ times the 
ground state energy of $-\nabla^2+V$.
\medskip

\textit{Step 4 (Minimizing in each box):}
Dropping the subsidiary condition $\suli n_\al=N$ can only lower the infimum. 
Hence it is sufficient to minimize each $Q_\al$ separately.
To justify the use of \eqref{basic}, we have to ensure that $n_\al$ is large 
enough. But if
the minimum is taken for some $\bar n_\al$, we have
\begin{equation}
\frac{\rmin}{\rmax}
\left(E^{\rm hom}(\bar n_\al+1,L)-E^{\rm hom}(\bar n_\al,L)\right)\geq 
8\pi a\rmax,
\end{equation}
and using Theorem \ref{chempot}, which states that
\begin{equation}
E^{\rm hom}(\bar n_\al+1,L)-E^{\rm hom}(\bar n_\al,L)\leq 8\pi a\frac{\bar 
n_\al}{L^3}
\left(1+O(\bar n_\al a^3/L^3)\right),
\end{equation}
we see that $\bar n_\al$ is at least $\sim \rmax L^3$. We shall later choose 
$L\sim N^{-1/10}$, so
the conditions needed for \eqref{basic} are fulfilled for $N$ large enough, 
since $\rmax\sim N$ and hence
$\bar n_\al\sim N^{7/10}$, $L/a\sim N^{9/10}$ and
$Y_\al\sim N^{-2}$. Thus we have (for large enough $N$)
\begin{equation}\label{qalpha}
Q_\al(f)\geq 4\pi 
a\left(\frac{\rmin}{\rmax}\frac{n_\al^2}{L^3}\left(1-CY_\al^{1/17}\right)
-2n_\al\rmax\right).
\end{equation}
We now use $Y_\al=a^3 n_\al /L^3\leq a^3 N/L^3\equiv Y$, and drop the 
requirement that $n_\al$ has to 
be an integer. The minimum of \eqref{qalpha} is obtained for
\begin{equation}
n_\al= \frac{\rmax^2}{\rmin}\frac{L^3}{(1-CY^{1/17})}.
\end{equation}
This gives for \eqref{ener2}
\begin{equation}
\begin{split}
&E^{\rm QM}(N,a)-E_R^{\rm GP}(N,a)\geq \\
&4\pi a\bar\rho_R N-4\pi a\suli_\al \rmin^2
L^3\left(\frac{\rmax^3}{\rmin^3}\frac{1}{(1-CY^{1/17})}\right).
\end{split}
\end{equation}
Now $\rho_R^{\rm GP}$ is differentiable by Lemma \ref{diff}, and strictly 
positive. Since  all the boxes are in the fixed cube $\Lambda_R$ there are 
constants 
$C'<\infty$, $C''>0$,
such that
\begin{equation}
\rmax-\rmin\leq NC'L,\quad \rmin\geq NC''.
\end{equation}
Therefore we have, for $Y$ and $L$ small,
\begin{equation}
\frac{\rmax^3}{\rmin^3}\frac{1}{(1-CY^{1/17})}\leq 
1+DY^{1/17}+D'L
\end{equation}
with suitable constants $D$ and $D'$. Also,
\begin{equation}
4\pi a\suli_\al \rmin^2 L^3\leq 4\pi a\int (\rho_R^{\rm GP})^2\leq E_R^{\rm 
GP}(N,a),
\end{equation}
and hence
\begin{equation}\label{yd}
E^{\rm QM}(N,a)\geq E_R^{\rm GP}(N,a)\left(1-DY^{1/17}-D'L\right).
\end{equation}
As last step it remains to optimize the length $L$. Recall that  $Y=a^3N/L^3$ 
and $Na$
is fixed. The exponents of $N$ in both error terms in
\eqref{yd} are the same for
\begin{equation}
L\sim aN^{9/10}\sim N^{-1/10}.
\end{equation}
The final result, therefore, is
\begin{equation}
E^{\rm QM}(N,a)\geq E_R^{\rm GP}(N,a)\left(1-D''N^{-1/10}\right).
\end{equation}
\end{proof}

\subsection{The limit theorems}

By Theorems \ref{upperthm} and \ref{lowerthm} we have (with $a=a_1/N$)
\begin{equation}
E^{\rm GP}(N,a)\left(1+O(N^{-2/3})\right)\geq E^{\rm QM}(N,a)\geq 
E_R^{\rm GP}(N,a)\left(1-O(N^{-1/10})\right).
\end{equation}
Dividing by $N$ and taking the limit $N\to\infty$ we have
\begin{equation}
E^{\rm GP}(1,a_1)\geq \lim_{N\to\infty}\frac{E^{\rm QM}(N,a_1/N)}{N}\geq
E_R^{\rm GP}(1,a_1)
\end{equation}
for all $R$ large enough. Using Lemma \ref{gpr} and taking the limit 
$R\to\infty$
we finally prove Theorem \ref{thm1}.

The convergence of the energy implies also the convergence of the densities: We 
replace $\V$ by $\V+\delta W$ with $W\in L^\infty$, and denote the corresponding
energies by $E_\delta(N,a)$. It is no restriction to 
assume that $V+\delta W\geq 0$ for  small $|\delta|$. 
$E^{\rm QM}_\delta(N,a_1/N)/N$ is concave in 
$\delta$ (it is an infimum over linear functions), and
converges for each $\delta$ to $E_\delta^{\rm GP}(1,a_1)$ as
$N\to\infty$.
This implies convergence of the derivatives and we have
(Feynman-Hellmann principle)
\begin{equation}
\left.\frac{\pd}{\pd\delta}E^{\rm QM}_\delta(N,a)\right|_{\delta=0}=\int_{\R^3} 
W\rho^{\rm QM}_{N,a},\quad
\left. \frac{\pd}{\pd\delta}E^{\rm GP}_\delta(N,a)\right|_{\delta=0}=\int_{\R^3}
W\rho^{\rm GP}_{N,a}
\end{equation}
with $\rho^{\rm QM}_{N,a}$ given by \eqref{qmdensity}. 
In the weak $L^1$ sense we thus have
\begin{equation}
\lim_{N\to\infty}\frac{1}{N}\rho^{\rm QM}_{N,a_1/N}(x)=\rho^{\rm GP}_{1,a_1}(x),
\end{equation}
which proves Theorem \ref{thm2}.

\section{Conclusions}

We have proved that the GP energy functional correctly 
describes the energy and particle 
density of a Bose gas in a trap to leading order in the small 
parameter $\bar\rho a^3$ (where $\bar\rho$ is the mean density and $a$
is the 
scattering length) in the limit where the particle number $N$ tends to infinity, 
but $a$ tends to zero with $Na$ fixed.

\appendix
\newcounter{zahler}
\renewcommand{\thesection}{\Alph{zahler}}
\renewcommand{\theequation}{\Alph{zahler}.\arabic{equation}}
\setcounter{zahler}{1}
\setcounter{equation}{0}
\section*{Appendix A}
In this appendix we prove Theorem  \ref{GPtheo}. The proof is split into several 
lemmas.

\begin{lem}[Strict convexity]\label{konv}
For $\rho\geq 0$, $\sqrt{\rho}\in\D$, $\E^{\rm GP}[\sqrt{\rho}]$ is strictly 
convex in
$\rho$.
\end{lem}
\begin{proof}
The second term in \eqref{gpfunct} is linear, the third quadratic in $\rho$. 
So it
suffices to show that the first term is convex. Let $\rho_1$ and $\rho_2$ be 
given, with $\Phi_1=\rho_1^{1/2}$ and
$\Phi_2=\rho_2^{1/2}$ in $\D_N$. Then also
$\Phi=\left(\alpha\rho_1+(1-\alpha)\rho_2\right)^{1/2}\in\D_N$
for all $0<\alpha<1$. We have
\begin{displaymath}
\begin{split}
\Phi\nabla\Phi &=\alpha\Phi_1\nabla\Phi_1+(1-\alpha)\Phi_2\nabla\Phi_2\\
&=\left(\alpha^{1/2}\Phi_1\right)\left(\alpha^{1/2}\nabla\Phi_1\right)+\left(
(1-\alpha)^{1/2}
\Phi_2\right)\left((1-\alpha)^{1/2}\nabla\Phi_2\right)\\
&\leq\left(\alpha\Phi_1^2+(1-\alpha)\Phi_2^2\right)^{1/2}
\left(\alpha|\nabla\Phi
_1|^2
+(1-\alpha)|\nabla\Phi_2|^2\right)^{1/2}\\
&=\Phi \left(\alpha|\nabla\Phi_1|^2
+(1-\alpha)|\nabla\Phi_2|^2\right)^{1/2}.
\end{split}
\end{displaymath}
Hence
\begin{displaymath}
|\nabla\Phi|^2\leq\alpha|\nabla\Phi_1|^2+(1-\alpha)|\nabla\Phi_2|^2.
\end{displaymath}
\end{proof}

\noindent
{\bf Remark.} 
Because $\V\geq 0$, $\E^{\rm GP}[\Phi]$ is also convex in $\Phi\in \D$. But 
since the domain $\D_N$ is 
not convex, it is necessary to consider $\rho\mapsto\E^{\rm GP}[\sqrt \rho]$.

\begin{lem}[Minimizer]\label{minimizer}
For all $N$ there exists a minimizing $\Phi_\infty\in\D_N$, with $\E^{\rm 
GP}[\Phi_\infty]=E^{\rm GP}(N,a)$. Moreover, $|\Phi_\infty|^2$ is unique.
\end{lem}
\begin{proof}
Let $\Phi_n$ be a minimizing sequence in $\D_N$, i.e., 
$\lim_{n\to\infty}\E^{\rm GP}[\Phi_n]=E^{\rm GP}$.
It is clear that there exists a constant $C$, such that 
$\|\nabla\Phi_n\|_2<C$,
$\|\Phi_n\|_4<C$ and $\int |\Phi_n|^2V<C$ for all $n$ (recall that $V$
is nonnegative). Hence the sequence belongs to a weakly 
compact set  in $L^4$, as well as in the Sobolev space $H^1=\{\Phi: 
\Vert\Phi\Vert_2^2+\Vert\nabla\Phi\Vert^2_2<\infty\}$, and in the
space $L^2_V$, defined by the $L^2$ norm 
$\Vert\Phi\Vert_V=(\int|\Phi(x)|^2V(x)d^3x)^{1/2}$. 
Thus, there exists a
$\Phi_\infty\in\D$ and a weakly 
convergent
subsequence, again denoted by $\Phi_n$, such that
\begin{gather}\nonumber
\Phi_n\rightharpoonup\Phi_\infty \text{ in } L^2\cap L^4\cap L^2_V\\ \nonumber
\nabla\Phi_n\rightharpoonup\nabla\Phi_\infty \text{ in } L^2.
\end{gather}

Because the $L^4$ norm, the Sobolev norm, and the $L^2_V$ norm  are all
weakly lower semicontinuous, we have 
\begin{displaymath}
\liminf_{n\to\infty}\E^{\rm GP}[\Phi_n]\geq\E^{\rm GP}[\Phi_\infty],
\end{displaymath}
and it remains only to
show that $\Phi_\infty\in\D_N$. Since $|\Phi_n|^2$ converges to 
$|\Phi_\infty|^2$ in 
$L^1_{\rm loc}$ it is clear that
$\|\Phi_\infty\|_2^2\leq N$. Moreover,
\begin{displaymath}
\int_B |\Phi_n|^2\xrightarrow{n\rightarrow\infty}\int_B |\Phi_\infty|^2\leq 
\|\Phi_\infty\|_2^2
\end{displaymath} for all bounded regions $B$. If $\|\Phi_\infty\|_2^2=N-\eps$ with 
$\eps>0$, then
there exists  a constant $M_B$ for all $B$, such that 
\begin{displaymath}
\int_{\R^3\backslash B}|\Phi_n|^2\geq\eps 
\end{displaymath}
for all $n>M_B$. Since $\lim_{|x|\to\infty}V(x)=\infty$, this would imply  
$\int\V|\Phi_n|^2\to\infty$, which is impossible because 
$\Phi_n$
is a minimizing sequence for the functional $\E^{\rm GP}$. Hence
$\|\Phi_\infty\|_2^2=N$.

The uniqueness of $|\Phi_\infty|^2$ follows immediately from strict convexity, 
Lemma 
\ref{konv}.
\end{proof}

\begin{lem}[GP equation]
Every minimizing $\Phi_\infty$ satisfies the Gross-Pitaevskii equation
\eqref{gpglg}. Conversely, every solution to \eqref{gpglg}, with $\mu$ given 
by
\eqref{mu}, is a minimizer for $\E^{\rm GP}$.
\end{lem}
\begin{proof}
Pick a function $f\in C_0^\infty$. The stationarity of $\E^{\rm GP}$ at 
$\Phi_\infty$ 
implies
\begin{displaymath}
\left.\frac{\pd}{\pd\eps}\left(\E^{\rm GP}[\Phi_\infty+\eps f]+\mu\|\Phi_\infty+\eps
f\|_2^2\right)\right|_{\eps=0}=0
\end{displaymath}
with a Lagrange parameter $\mu$ to take account of the subsidiary condition 
$\Vert \Phi\Vert_2^2=N$. With $f$ real valued one obtains
\begin{displaymath}
-\nabla^2\real\,\Phi_\infty+V\real\,\Phi_\infty+8\pi
a|\Phi_\infty|^2\real\,
\Phi_\infty=\mu\real\,\Phi_\infty
\end{displaymath}
and an analogous equation for $\imag\,\Phi_\infty$ with $f$ purely imaginary. The 
value 
of $\mu$ is obtained by multiplying the GP equation with $\Phi_\infty$ and 
integrating. By the same argument, every solution $\Phi$ to the GP equation 
satisfies 
$\E^{\rm GP}[\Phi]=E^{\rm GP}$ and is thus a minimizer.
\end{proof}

\begin{lem}[Uniqueness]\label{phase}
The minimizing $\Phi_\infty$ is unique up to a constant phase factor. 
This factor can be chosen so that $\Phi_\infty$ is strictly positive.
\end{lem}
\begin{proof}
Since  $\E^{\rm GP}[|\Phi|]\leq\E^{\rm GP}[\Phi]$ (by an analogous computation
as in the proof of Lemma 
\ref{konv}), we know that $|\Phi_\infty|$ is a minimizer and hence a solution to 
the 
GP equation. It is thus an eigenstate of the  Hamiltonian
$H=-\nabla^2+W$
with $W=\V+8\pi a|\Phi_\infty|^2$ (recall that $|\Phi_\infty|^2$ is unique), and since 
it 
is nonnegative, it must be a ground state. Since $\Phi_\infty$ solves the same 
equation it is also a ground state. Now $W\in L^2_{\rm loc}$ and
$\lim_{|x|\to\infty}W(x)=\infty$,
so the ground state of $H$ is unique up to a phase and without zeros (see
\cite{reed}, XIII.47).  
\end{proof}
\medskip

The unique strictly positive minimizer is denoted by $\Phi^{\rm GP}$.

\begin{lem}[Exponential fall-off]
For all $t>0$ there exists an $M_t$, such that $\Phi^{\rm GP}(x)\leq M_t 
e^{-t|x|}$. 
In particular,
$\Phi^{\rm GP}\in L^\infty$.
\end{lem}
\begin{proof}
Put $W=\V+8\pi a(\Phi^{\rm GP})^2$ and let $t>0$. The GP equation implies
\begin{displaymath}
\left(-\nabla^2+t^2\right)\Phi^{\rm GP}=-\left(W-\mu-t^2\right)\Phi^{\rm GP}.
\end{displaymath}
Using the Yukawa potential $Y_t(x)=(4\pi|x|)^{-1}\exp(-t|x|)$ we can
rewrite this as
\begin{displaymath}
\Phi^{\rm GP}(x)=-\int Y_t(x-y)(W(y)-\mu-t^2)\Phi^{\rm GP}(y)d^3y.
\end{displaymath}
Since $\Phi^{\rm GP}>0$, and $W(y)-\mu-t^2>0$ for $|y|>R$ with $R$ large 
enough, 
we also have
\begin{displaymath}
\Phi^{\rm GP}(x)\leq -\inli_{|y|<R} Y_t(x-y)(W(y)-\mu-t^2)\Phi^{\rm
GP}(y)d^3y.
\end{displaymath}
Now $W\Phi^{\rm GP}\in L^2_{\rm loc}$, and hence
\begin{displaymath}
M_t=\sup_x \inli_{|y|<R}\frac{\exp 
\{t(|x|-|x-y|)\}}{4\pi|x-y|}(W(y)-\mu-t^2)\Phi^{\rm 
GP}(y)d^3y<\infty.
\end{displaymath}
\end{proof}

\begin{lem}[Regularity]\label{diff}
 $\Phi^{\rm GP}(x)$ is once continuously differentiable in $x\in\R^3$, and $\nabla\Phi^{\rm GP}$ 
is H\"older continuous of order $1$. If $\V\in C^\infty$, then
\mbox{$\Phi^{\rm GP}\in C^\infty$}. Moreover, $E^{\rm GP}(N,a)$ is
continuously differentiable in $a$ and hence in $N$ (by Eq.\
\eqref{scaling}), and $dE^{\rm GP}(N,a)/dN$ satisfies \eqref{mu}.
\end{lem}
\begin{proof}
The last lemma and the GP equation imply $\nabla^2\Phi^{\rm GP}\in 
L^\infty_{\rm loc}$. Thus $\nabla\Phi^{\rm GP}$ exists and is H\"older 
continuous 
(see \cite{lieb}, 
10.2). The
$C^\infty$ property follows by a bootstrap argument. The
differentiability with respect to the parameter $a$ may be shown by a 
Feynman-Hellmann type
argument like analogous statements (e.g. differentiability w.r.t.\
nuclear charges) in TF theory \cite{lieb81}. Eq.\
\eqref{mu}
follows immediately from \eqref{scaling} and  $E^{\rm GP}(1,Na)=
\E^{\rm GP}[\Phi^{\rm GP}_{1,Na}]$.
\end{proof}
Lemmas \ref{konv}--\ref{diff} complete the proof of Theorem \ref{GPtheo}.  
\medskip

\addtocounter{zahler}{1}
\setcounter{equation}{0}
\section*{Appendix B}
In this appendix we show that \eqref{basic} holds for nonnegative potentials 
satisfying \eqref{condition}, and that a similar estimate with $1/17$ replaced 
by $O(\varepsilon)$ holds under the condition \eqref{condition2}.

We cut the potential at a finite radius
$\RR$ which, because of $v\geq 0$, can only decrease the energy. We thus define
\begin{equation}
\vv(r)=v(r)\Theta(\RR-r)
\end{equation}
and denote the corresponding scattering length by $\as\leq a$. Let $u$ be the 
zero energy scattering solution for the potential $v$ (cf.\ \eqref{scatteq})
and put
\begin{equation}
h(r)=r-\frac{u(r)}{u'(r)}.
\end{equation}
The difference  $a-\as$ can be estimated as follows. Since  $v(r)$ 
and  $\widetilde v(r)$ agree for $r\leq \widetilde R$, the same holds
for the corresponding scattering solutions. Moreover,   $\widetilde
a=h(\widetilde R)$. Hence 
\begin{equation}   
\begin{split}\label{absch}
a-\as&=\inli_{\RR}^\infty h'(r)dr=\inli_{\RR}^\infty 
\frac{u(r)u''(r)}{u'(r)^2}dr\\
&\leq\inli_{\RR}^\infty \frac{u''(r)}{u(r)}r^2dr=\half\inli_{\RR}^\infty 
v(r)r^2dr,
\end{split}
\end{equation}
where convexity of $u$ has been used to derive the inequality. We remark that for  
$\RR\to 0$ this simple estimate gives the
\textit{Spruch-Rosenberg} inequality \cite{sr}
\begin{equation}
a\leq\half\inli_0^\infty v(r)r^2dr.
\end{equation}
Assuming \eqref{condition}
one obtains
\begin{equation}\label{fehler}
\as\geq a\left(1-{\rm 
const.\,}\left(\frac{a}{\RR}\right)^{\frac{1}{5}+\eps}\right).
\end{equation}
Eq.\ \eqref{basic} holds in any case with $a$ replaced by $\widetilde
a$, and if we we take
$\RR\propto a Y^{-5/17+\eps'}$
with $\epsilon'>0$  then the error in \eqref{fehler} is of 
higher order than the leading error term in
\eqref{basic}. We have thus established \eqref{basic} under the condition 
\eqref{condition}. If only the weaker condition \eqref{condition2} holds, then
the additional error term may be $O(Y^{({5}/{17}-\epsilon')\epsilon})$.

To see the significance of condition \eqref{condition2} we also estimate 
$a-\as$ from below: 
\begin{equation}
a-\as\geq \inli_{\RR}^\infty u(r)u''(r)\geq \half\inli_{\max(\RR,a)}^\infty
v(r)(r-a)^2dr.
\end{equation}
In order that $a$ is finite the last integral must converge, i.e.,  a slower 
decrease than $1/r^3$ is not allowed.


\begin{thebibliography}{99}
\bibitem{G1961}E.P. Gross, {\it Structure of a Quantized Vortex in
    Boson Systems,} Nuovo Cimento {\bf 20}, 454--466 (1961)
\bibitem{P1961} L.P. Pitaevskii, {\it Vortex lines in an imperfect
    Bose gas,} Sov. Phys. JETP, {\bf 13}, 451--454 (1961)

\bibitem{G1963} E.P. Gross, {\it Hydrodynamics of a superfluid
condensate,} J. Math. Phys. {\bf 4}, 195--207 (1963)
\bibitem{DGPS}
F. Dalfovo, S.\ Giorgini, L.P.\ Pitaevskii, S.\ Stringari, {\it Theory of 
Bose-Einstein condensation in trapped gases},
Rev. Mod. Phys. \textbf{71}, 
463--512 (1999)
\bibitem{LY1998}
E.H. Lieb, J. Yngvason, {\it Ground State Energy of the low density Bose Gas},
Phys. Rev. Lett. \textbf{80}, 2504--2507 (1998)
\bibitem{dyson}
F.J. Dyson, {\it Ground-State Energy of a Hard-Sphere Gas}, 
Phys. Rev. \textbf{106}, 20--24 (1957)
\bibitem{reed}
M. Reed, B. Simon, {\it Methods of Modern Mathematical Physics IV}, Academic Press, 
1978
\bibitem{lieb81}E.H. Lieb, {\it Thomas-fermi and related theories of atoms and 
molecules}, Rev. Mod. Phys. {\bf 53}, 603--641 (1981)
\bibitem{lieb}
E.H. Lieb, M. Loss, {\it Analysis}, Amer. Math. Society, 1997
\bibitem{lieb63} E.H. Lieb, {\it 
Simplified Approach to the Ground State Energy of an Imperfect Bose Gas}, 
Phys. Rev. {\bf 130}, 2518--2528 (1963). See also Phys. Rev. {\bf 133},
A899-A906 (1964) (with A.Y. Sakakura) and Phys. Rev. {\bf 134},
A312-A315 (1964) (with W. Liniger).

\bibitem{lions}
P.L. Lions, {\it Two Geometrical Properties of Solutions of Semilinear 
Problems}, Applicable Analysis, Vol. 12, 267-272 (1981)
\bibitem{sr}
L.~Spruch, L.~Rosenberg, {\it Upper bounds on Scattering Lengths for
Static Potentials}, Phys. Rev. \textbf{116}, 1034 (1959)
\end{thebibliography}
\end{document}